\documentclass[10pt,twocolumn,prl,aps,amssymb,amsmath,showpacs,tightenlines]{revtex4-1}
\usepackage{dsfont}
\usepackage[dvips]{graphicx}
\usepackage{color}
\usepackage{subfigure}

\newcommand{\re}{\mbox{$\rm e$}}
\newcommand{\ri}{\mbox{$\rm i$}}
\newcommand{\rd}{\mbox{$\rm d$}}

\begin{document}

\title{Mixed-state evolution in the presence of gain and loss}

\author{Dorje~C.~Brody${}^1$ and Eva-Maria~Graefe${}^2$}

\affiliation{${}^1$Mathematical Sciences, Brunel University, Uxbridge UB8 3PH, UK \\ 
${}^2$Department of Mathematics, Imperial College London, London SW7 2AZ, UK }

%\date{\today}

\begin{abstract}
A model is proposed that describes the evolution of a mixed state of a quantum system 
for which gain and loss of energy or amplitude are present. Properties of the 
model are worked out in detail. In particular, invariant subspaces of the space of density 
matrices corresponding to the fixed points of the dynamics are identified, and the 
existence of a transition between the phase in which gain and loss are balanced and the 
phase in which this balance is lost is illustrated in terms of the time average of observables. 
The model is extended to include a noise term that results from a uniform random perturbation 
generated by white noise. Numerical studies of example systems show the emergence of 
equilibrium states that suppress the phase transition. 
\end{abstract}

\pacs{03.65.Ca, 05.30.Rt, 03.65.Yz}

\maketitle

Over the past decade there have been considerable research interests, both 
theoretical and experimental, into the static and dynamic properties of classical 
and quantum systems for which gain and loss are present \cite{NM0,jpa}. This is in 
part motivated by the realisation that when a system is placed in a configuration 
in which its energy or amplitude is transferred into its environment through one 
channel, but at the same time is amplified by the same amount through another 
channel, the resulting dynamics can exhibit features that are similar to those 
seen in Hamiltonian dynamical systems. The time evolution of such a system 
can be described by a Hamiltonian that is symmetric under a space-time 
reflection, that is, invariant under the parity-time (PT) reversal. 

Interest in the theoretical study of PT symmetry was triggered by the 
discovery that complex PT-symmetric quantum Hamiltonians can possess entirely 
real eigenvalues \cite{BB}. One distinguishing feature of PT-symmetric quantum 
systems is the existence of phase transitions associated with the breakdown of the 
symmetry. That is, depending on the values of the matrix elements of the Hamiltonian, 
its eigenstates may or may not be symmetric under the parity-time reversal. In the 
`unbroken phase' where the eigenstates respect PT symmetry, the eigenvalues 
are real and there exists a similarity transformation that maps a local 
PT-symmetric Hamiltonian into a typically nonlocal Hermitian Hamiltonian 
\cite{Mostafa,BBJ}; whereas in the `broken phase' the eigenvalues constitute complex 
conjugate pairs. The transition between these phases is similar to second-order phase 
transitions in statistical mechanics, accompanied by singularities in the covariance 
matrix of the estimators for the parameters in the Hamiltonian (this can be seen by 
using methods of information geometry \cite{BH}). 

Experimental realisations of these phenomena are motivated in part by the observation 
that the presence of a loss, which traditionally has been viewed as an undesirable feature, 
can positively be manipulated so as to generate unexpected interesting effects 
(cf. \cite{DC1}). In particular, PT phase transitions have been predicted or observed in 
laboratory experiments for a range of systems; most notably in optical 
waveguides \cite{NM,DC2,DC3}, but also in laser physics \cite{DS,SR,SR2}, 
in electric circuits \cite{TK,TK2}, or in microwave cavity \cite{BD}. 

In the quantum context, theoretical investigations into the properties and predictions 
of the evolution equation where gain and loss are present have thus far been confined 
primarily to pure states. However, quantum systems are commonly described by 
mixed states, especially when they are subject to manipulations. Therefore, to describe 
or predict the behaviour of such quantum systems it is necessary to understand how 
a given mixed state might evolve. For this purpose we seek a complex extension of the 
von-Neumann equation that reduces to the Schr\"odinger equation when restricted to 
pure states. 

To describe various forms of dissipation or noise, Lindblad-type equations \cite{AK,GL} 
are often considered. They can be related to pure state evolutions with complex 
Hamiltonians, but typically involve stochastic terms \cite{PK} that distinguish 
them from the complex Schr\"odinger equation. In particular, an initially pure state tends 
to evolve into a mixed state due to the influence of noise. Here we propose an alternative 
model that describes the evolution of a generic density matrix in the context of gain and 
loss such that it reduces to the complex Schr\"odinger equation for pure states. 
This model can be further extended to include additional dissipation or noise in the form 
of Lindblad terms or other effective descriptions, depending on the context. Our model is 
given by the dynamical equation 
\begin{eqnarray}
\frac{\rd \rho}{\rd t} = -\ri [H,\rho] - \big( [{\mathit\Gamma},\rho]_+ - 2\, {\rm tr}
(\rho{\mathit\Gamma}) \rho \big), 
\label{eq:1}
\end{eqnarray}
where $H=H^\dagger$ is the Hermitian part of the Hamiltonian generating ambient 
unitary motion, ${\mathit\Gamma}={\mathit\Gamma}^\dagger$ is the skew-Hermitian 
part of the Hamiltonian governing gain and loss, and $[{\mathit\Gamma},\rho]_+ = 
{\mathit\Gamma}\rho+\rho{\mathit\Gamma}$ denotes the symmetric product. We 
refer to (\ref{eq:1}) as the \textit{covariance equation}, on account of the structure of 
the term involving ${\mathit\Gamma}$. In this Letter we shall motivate  
equation (\ref{eq:1}), investigate its properties in detail, identify the structure of 
stationary states, and show that a PT phase transition manifests itself by means 
of the time average of observables. 

An experimental implementation of an evolution of a purely quantum 
system requires a careful balancing of loss and gain. It should be evident, however, 
that the realisation of such an evolution for a pure state is difficult, to say the least, 
since energetic manipulations of a quantum system inadvertently perturb the system. 
To account for the possible impact of uncontrollable ambient noise, we consider an 
extension of the model (\ref{eq:1}) that includes an additional term generated by 
uniform Gaussian perturbations. We shall study dynamical aspects of the extended 
model, identify the existence of equilibrium states by means of numerical studies, 
and show how PT phase transitions can be suppressed by noise. 

Let us begin by motivating the introduction of the covariance equation (\ref{eq:1}). 
In the case of pure states, the norm-preserving evolution equation generated by a 
complex Hamiltonian $K=H-\ri{\mathit\Gamma}$ is given by 
\begin{eqnarray}
\frac{\rd|\psi\rangle}{\rd t} = - \ri (H-\langle H\rangle)|\psi\rangle - ({\mathit\Gamma}- 
\langle{\mathit\Gamma}\rangle) |\psi\rangle . 
\label{eq:2}
\end{eqnarray}
Together with an additional equation for the overall probability ${\dot N}=-
\langle {\mathit\Gamma}\rangle N$, equation (\ref{eq:2}) is equivalent to the familiar 
complexified Schr\"odinger equation. The benefit of (\ref{eq:2}) is that it is 
defined on the projective Hilbert space. Besides the covariance equation 
(\ref{eq:1}), there are many alternative evolution equations for 
mixed states that reduce to (\ref{eq:2}) when restricted to a pure state satisfying 
$\rho^2=\rho$. An example is given by the 
double-bracket equation 
\begin{eqnarray}
\frac{\rd\rho}{\rd t} = -\ri [H,\rho] - [[{\mathit\Gamma} ,\rho], \rho] . 
\label{eq:3}
\end{eqnarray}
For ${\mathit\Gamma} \propto H$, the norm-preserving equations (\ref{eq:2}) 
and (\ref{eq:3}) have been considered by Gisin \cite{Gis} as candidate 
equations to describe dissipative quantum evolution. That (\ref{eq:1}) and 
(\ref{eq:3}) are identical for a pure state can be seen by observing that 
$\rho{\mathit\Gamma}\rho={\rm tr}(\rho{\mathit\Gamma}) \rho$ when $\rho^2=\rho$.
Among the possible dynamical equations, the covariance equation (\ref{eq:1}) is singled 
out on account of the fact that its formal solution, given an initial state $\rho_0$, can be 
expressed in the form 
\begin{eqnarray}
\rho_t = \frac{\re^{-{\rm i}(H-{\rm i}{\mathit\Gamma})t} \rho_0 
\re^{{\rm i}(H+{\rm i}{\mathit\Gamma})t}} 
{{\rm tr} \big({\re^{-{\rm i}(H-{\rm i}{\mathit\Gamma})t} \rho_0 
\re^{{\rm i}(H+{\rm i}{\mathit\Gamma})t}}\big)} , 
\label{eq:5} 
\end{eqnarray} 
which provides a natural generalisation of its unitary counterpart when 
${\mathit\Gamma}=0$. Similarly, the dynamical equation satisfied by an observable 
$\langle F\rangle={\rm tr}(F \rho_t)$ reads 
\begin{eqnarray}
\frac{\rd \langle F\rangle}{\rd t} = \ri \langle[H,F]\rangle - \langle[{\mathit\Gamma},F]_+
\rangle + 2\, \langle{\mathit\Gamma}\rangle \langle F\rangle , 
\label{eq:4}
\end{eqnarray}
which agrees with the complex extension of the Heisenberg equation of motion 
obtained in Refs.~\cite{ref,GKN,GHK} for pure states. Note that the 
covariance-type structure in (\ref{eq:1}) has also appeared in the contexts of approach 
to thermal equilibrium \cite{KS}, dissipative motion \cite{mdo,Sergi}, and constrained 
quantum dynamics \cite{BGH}. 

Key properties of the evolution equation (\ref{eq:1}) can be summarised as 
follows: (i) It preserves the overall probability so that ${\rm tr}(\rho_t)=1$ for all $t\geq0$. 
This can be checked by verifying the relation $\partial_t {\rm tr}(\rho)=0$. (ii) Unlike a 
unitary time evolution, it does not in general preserve the purity of the state. In particular, 
we have  
\begin{eqnarray}
\frac{\rd}{\rd t}\, {\rm tr}\,\rho^2 = -4 \big( {\rm tr}({\mathit\Gamma}\rho^2)  
- {\rm tr}(\rho{\mathit\Gamma}) {\rm tr}(\rho^2) \big), 
\label{eq:6} 
\end{eqnarray}
and in general the right side of (\ref{eq:6}) is nonzero when $\rho^2\neq\rho$. Thus, the 
purity of the initial state is not preserved by (\ref{eq:1}) when $\rho$ is not a fixed point 
of the dynamics, but an initially pure state will remain pure. When $H=0$, we have 
the relation  
\begin{eqnarray}
\frac{\rd}{\rd t}\, {\rm tr}(\rho{\mathit\Gamma}) = -2\, {\rm var}_\rho({\mathit\Gamma}) 
\leq 0 ,
\label{eq:7} 
\end{eqnarray}
from which it follows that: (iii) The imaginary part ${\mathit\Gamma}$ of the Hamiltonian 
drives every state towards the ground 
state of ${\mathit\Gamma}$. (iv) The evolution equation (\ref{eq:1}) preserves the 
positivity of $\rho$, and is `autonomous' in 
the sense that the dynamical trajectory in the space of density matrices is determined 
uniquely by the specification of the initial density matrix $\rho_0$, and is not dependent 
on the kind of probabilistic mixture the initial state might represent. (v) In the case of a 
unitary motion, the speed $v={\rm tr}(\partial_t\sqrt{\rho})^2$ of the evolution of the state 
is constant of motion and is given by the Wigner-Yanase skew information 
$v=2{\rm tr}(H^2\rho) - 2{\rm tr}(H\sqrt{\rho}H\sqrt{\rho})$ \cite{DCB}, which reduces to the 
Anandan-Aharonov relation $v=2\Delta H^2$ for pure states $\sqrt{\rho}=\rho$. When the 
dynamics is governed by a complex Hamiltonian $H-\ri{\mathit\Gamma}$, the evolution 
speed is not a constant of motion and is given by the expression 
\begin{eqnarray}
v &=& 2 \left( {\rm tr}(H^2\rho) - {\rm tr}(H\sqrt{\rho}H\sqrt{\rho}) \right) - 2\ri \, 
{\rm tr}([H,{\mathit\Gamma}]\rho) \nonumber \\ && + 2 \left( {\rm tr}({\mathit\Gamma}^2\rho) 
+ {\rm tr}({\mathit\Gamma}\sqrt{\rho}{\mathit\Gamma}\sqrt{\rho}) -2 \big( {\rm tr}(
{\mathit\Gamma}\rho) \big)^2 \right),
\label{eq:8} 
\end{eqnarray}
which reduces to $v=2\Delta H^2 + 2\Delta {\mathit\Gamma}^2 - 2\ri \langle [{\mathit\Gamma},
H]\rangle$ for pure states. 

Next we identify the fixed points of the dynamics. We begin by the following elementary 
observation concerning eigenstates of a generic complex Hamiltonian $K=H - \ri 
{\mathit\Gamma}$. Suppose that $|\phi\rangle$ is a normalised eigenstate of $K$:
\begin{eqnarray}
K|\phi\rangle = \lambda |\phi\rangle . 
\label{eq:9} 
\end{eqnarray}
Then, if $\lambda=E-\ri \gamma$, where $E,\gamma$ real, we have 
$\langle\phi|H|\phi\rangle=E$ 
and $\langle\phi|{\mathit\Gamma}|\phi\rangle=\gamma$. In particular, an eigenfunction 
$|\phi\rangle$ of a complex Hamiltonian $K$ has a real eigenvalue if and only if 
$\langle\phi|{\mathit\Gamma}|\phi\rangle=0$. With this in mind, we establish the following 
results: (vi) Every eigenstate of the Hamiltonian $K$, irrespective of whether the 
eigenvalue is real, is a fixed point of the motion (\ref{eq:1}). They are the only 
stationary states that are pure. (vii) The mixed stationary states of the dynamical equation 
(\ref{eq:1}) consist of convex combinations of the eigenstates of the the Hamiltonian $K$ 
associated with \textit{real} eigenvalues. Furthermore, the totality of mixed stationary states 
lies on the subspace of density matrices for which ${\rm tr}({\mathit\Gamma}\rho)=0$. 

The statement (vi) can be established as follows: From (\ref{eq:5}), if we set $\rho_0=
|\phi\rangle\langle\phi|$, where $|\phi\rangle$ is an eigenstate of $K$ with eigenvalue 
$E-\ri\gamma$, then we have 
\begin{eqnarray}
\re^{-{\rm i}K t}\rho_0\re^{{\rm i}K^\dagger t}=
\re^{-2\gamma t}|\phi\rangle\langle\phi|, 
\end{eqnarray}
and hence ${\rm tr}(\re^{-{\rm i}K t}\rho_0\re^{{\rm i}K^\dagger t})=\re^{-2\gamma t}$. 
Putting these together, it follows that $\rho_t=\rho_0$. To establish (vii) we set 
\begin{eqnarray}
\rho_0 = \sum_{k=1}^m p_k |\phi_k\rangle\langle\phi_k| , 
\label{eq:11}
\end{eqnarray}
where $\{p_k\}$ are nonnegative numbers adding to unity, $\{|\phi_k\rangle\}$ are 
eigenstates of $K$ with real eigenvalues, and $m$ is the number of real eigenvalues.  
Substituting (\ref{eq:11}) in (\ref{eq:5}), a short calculation shows that $\rho_t=
\rho_0$. To show that the stationarity breaks down if the convex combination contains 
eigenstates with complex eigenvalues, suppose that we add terms in (\ref{eq:11}) 
associated with complex eigenvalues. Then from (\ref{eq:5}) we obtain 
\begin{eqnarray}
\rho_t = \frac{\sum_j p_j \re^{-2\gamma_j t}|\phi_j\rangle\langle\phi_j|}
{\sum_j p_j \re^{-2\gamma_j t}} ,
\label{eq:12}
\end{eqnarray}
where $\gamma_j=\langle\phi_j|{\mathit\Gamma}|\phi_j\rangle$. Since at least one of the 
$\{\gamma_j\}$ is nonzero by assumption, it follows that $\rho_t\neq \rho_0$. Finally, 
since all terms in (\ref{eq:11}) have the property that $\langle\phi_k|{\mathit\Gamma} 
|\phi_k\rangle=0$, it follows that ${\rm tr}({\mathit\Gamma}\rho)=0$ for all stationary 
states. 

A corollary to (\ref{eq:12}) is that: (viii) If  the initial state $\rho_0$ admits an eigenfunction 
expansion such that one or more terms are associated with complex eigenvalues, then 
\begin{eqnarray}
\lim_{t\to\infty} \rho_t = |\phi^*\rangle\langle\phi^*|, 
\label{eq:13}
\end{eqnarray}
where $|\phi^*\rangle$ is the member of the eigenfunctions $\{|\phi_j\rangle\}$ in the 
expansion for which the associated imaginary part $\gamma_j$ of the eigenvalue takes 
maximum value. Hence the flow structure in the space of density matrices, when there 
are imaginary eigenvalues, is rather complex and intricate in higher dimensions, where 
for each eigenstate $|\phi\rangle\langle\phi|$ associated with an imaginary eigenvalue 
there is a continuum of states for which $|\phi\rangle\langle\phi|$ is an asymptotic 
attractor. If, however, there is no real eigenvalue at all, then this segmentation disappears 
and the eigenstate with the largest $\gamma$ becomes the single attractor. 

To investigate the behaviour in the PT-symmetric phase where all eigenvalues are real, 
let us put 
\begin{eqnarray}
\rho_0 = \sum_{j,k} \rho_{jk} |\phi_j\rangle\langle\phi_k| . 
\label{eq:14}
\end{eqnarray}
Then the trace condition ${\rm tr}(\rho_0)=1$ implies that 
\begin{eqnarray}
\sum_{j=1}^m \rho_{jj} + \sum_{j\neq k} \rho_{jk} \langle\phi_k|\phi_j\rangle = 1,   
\label{eq:15}
\end{eqnarray}
since the eigenfunctions are not orthogonal when ${\mathit\Gamma}\neq0$. 
Substituting (\ref{eq:14}) in (\ref{eq:5}) we find 
\begin{eqnarray}
\rho_t = \frac{\sum_{j,k} \rho_{jk} \re^{-{\rm i}\omega_{jk}t}|\phi_j\rangle\langle\phi_k|}
{\sum_{j,k} \rho_{jk} \re^{-{\rm i}\omega_{jk}t}\langle\phi_k|\phi_j\rangle} ,
\label{eq:16}
\end{eqnarray}
where $\omega_{jk}=E_j-E_k$. We thus deduce the following: (ix) When all 
eigenvalues of the Hamiltonian $K$ are real, every orbit is periodic if the energy 
eigenvalues are commensurable; otherwise, the evolution is typically ergodic 
on a small toroidal subspace of the space of density matrices containing $\rho_0$. 
Therefore, in the PT-symmetric phase, dynamical features of the system are 
analogous to those of a unitary system, albeit differences in detail such as the 
lack of constancy of the evolution speed. The similarity is due to the fact that the 
nonlinearity of (\ref{eq:1}) is merely to preserve ${\rm tr}(\rho)$; hence the 
evolution equation cannot generate nontrivial fixed points such as saddle points. 

\begin{figure}[t]
\centering
\subfigure[~unbroken phase]{
\includegraphics[scale=0.3]{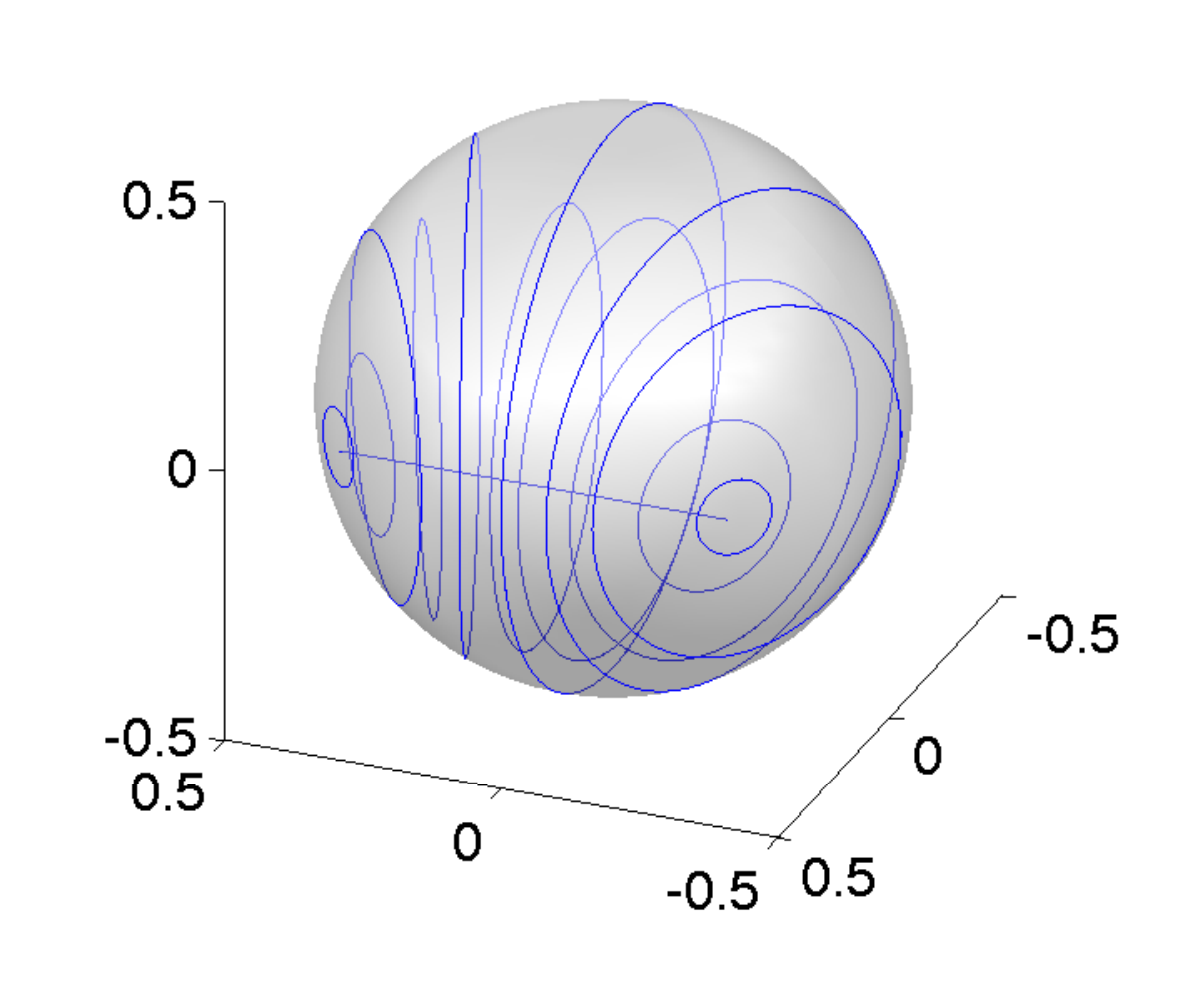}
\label{fig:1.1}}
\qquad
\subfigure[~broken phase]{
\includegraphics[scale=0.3]{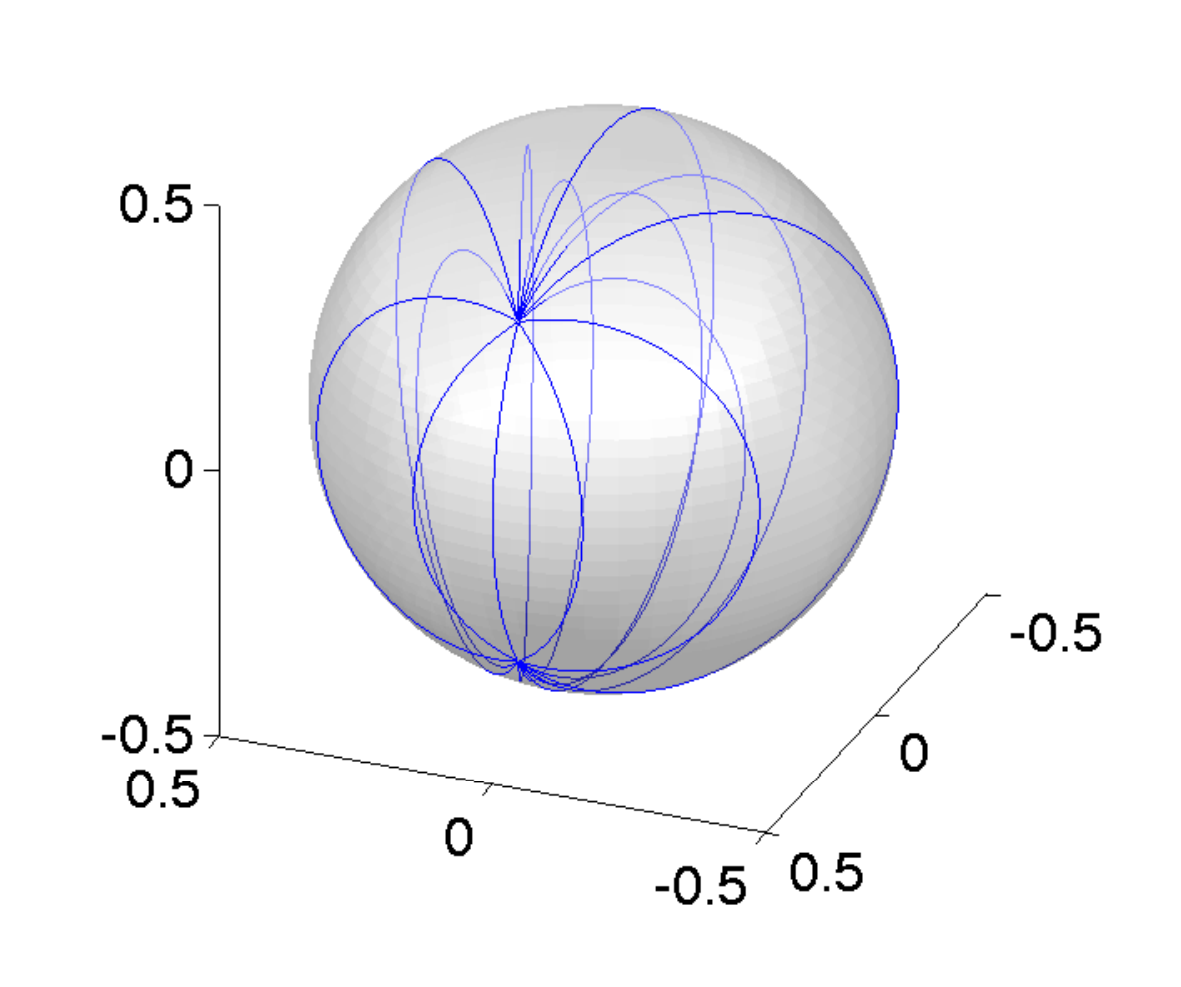}
\label{fig:1.2}}
\caption{\footnotesize (colour online) 
\textit{Trajectories generated by the Hamiltonian $K=\sigma_x - \ri \gamma 
\sigma_z$}. 
In the unbroken phase ($|\gamma|\leq1$) all orbits are closed, and the fixed-points 
are represented by a straight line. In the 
broken phase ($|\gamma|>1$) a sink and a source emerges and all states 
converge to the ground state in the limit $t\to\infty$.  
\label{fig:1}
}
\end{figure}

\begin{figure}[t]
\centering
\subfigure[~$\kappa = 0$]{
\includegraphics[scale=0.26]{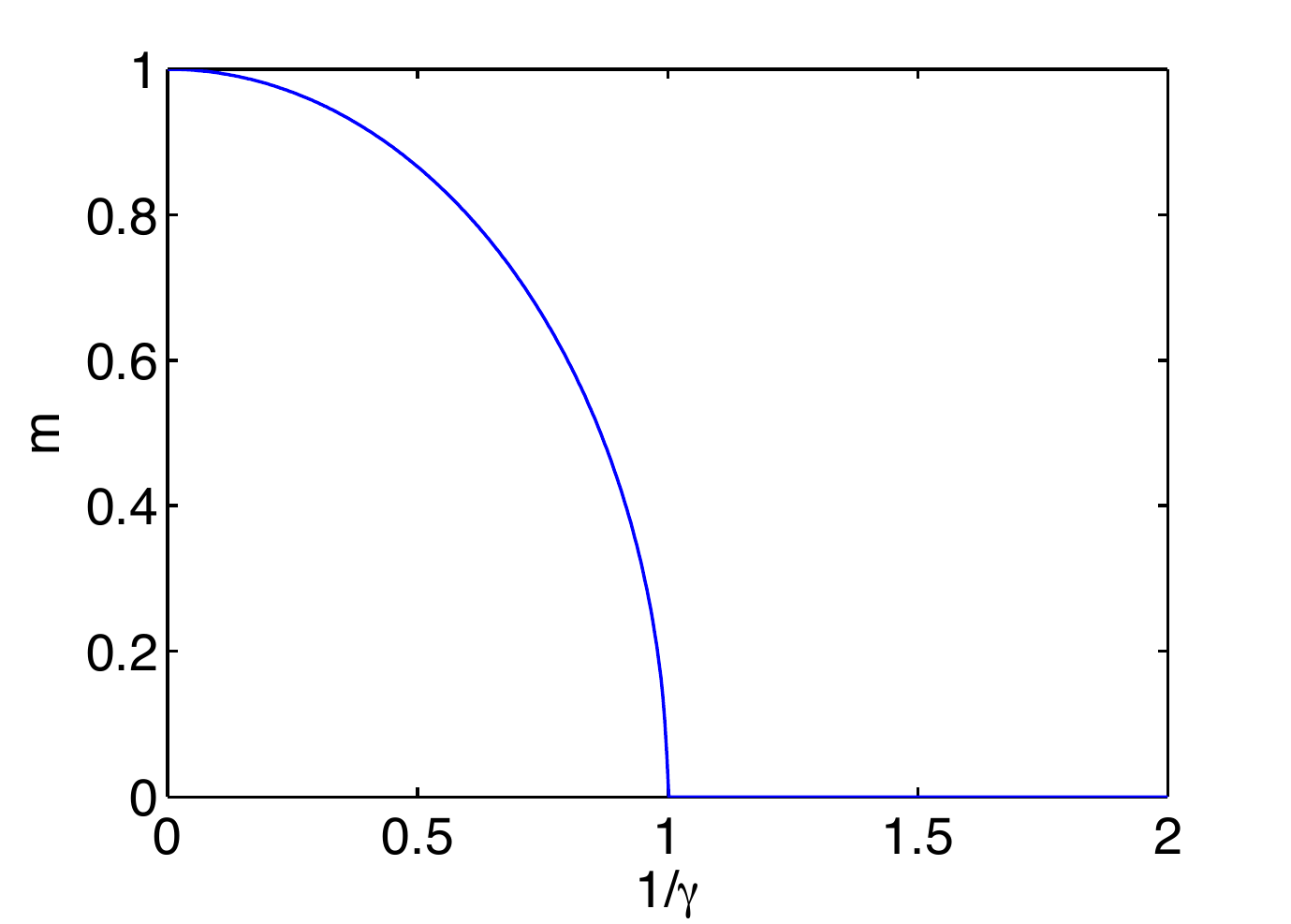}
\label{fig:2.1}}
\qquad
\subfigure[~$\kappa \neq 0$]{
\includegraphics[scale=0.26]{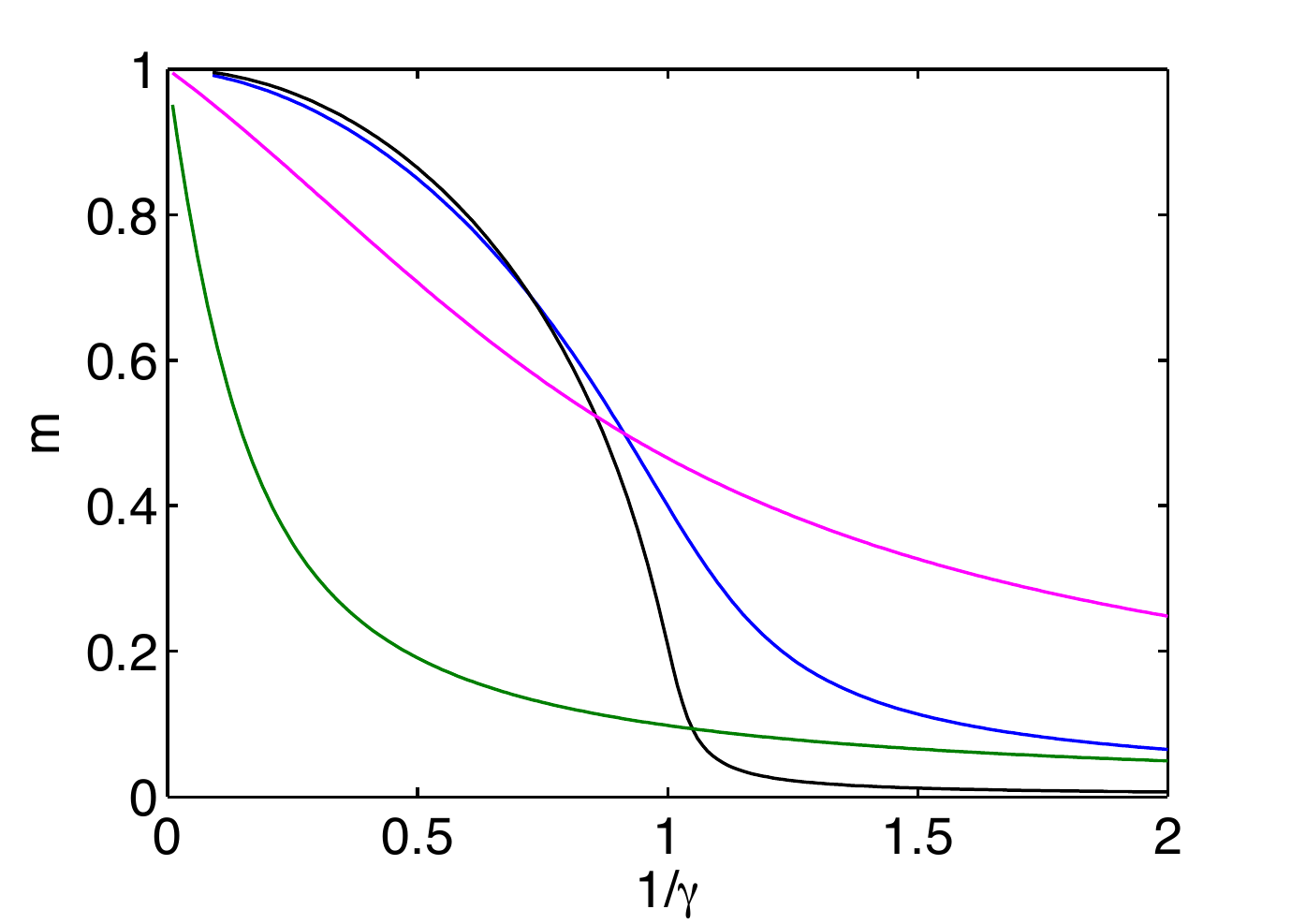}
\label{fig:2.2}}
\caption{\footnotesize (colour online) 
\textit{Time and ensemble averages of the observable $\sigma_z$ as functions of 
$\gamma^{-1}$}. 
(a) When $\kappa=0$, irrespective of the initial state $\rho_0$, the time average of 
${\rm tr}(\sigma_z \rho_t)$ shows the existence of a phase transition at the critical 
point $\gamma_c=1$. (b) When $\kappa\neq0$ an equilibrium state $\rho^*$ 
emerges for each value of $\gamma,\kappa$, to which every initial state $\rho_0$ 
converges. Hence a time average can be replaced by an ensemble average. For 
$\kappa>0$ the phase transition is eliminated, as indicated here for $\kappa=0.01$ 
(black), $0.1$ (blue), $1$ (magenta), and $10$ (green). 
\label{fig:2}
}
\end{figure}

In figure~\ref{fig:1} we illustrate integral curves of (\ref{eq:1}) for a two-level example 
system with the Hamiltonian $K=\sigma_x - \ri \gamma \sigma_z$, where $\gamma$ 
is a real parameter. Note that the eigenvalues of $K$ are given by $\lambda_\pm=
\pm\sqrt{1-\gamma^2}$. Hence the symmetry is broken if $|\gamma|>1$. Observe that 
in the unbroken phase the orbits are periodic, and hence the purity of mixed states 
oscillates, while in the broken phase every mixed state is asymptotically purified to the 
ground state. To 
visualise the effect of the phase transition it will be useful to introduce an `order 
parameter' $m$ by the time average of $\sigma_z$:
\begin{eqnarray}
m = \lim_{T\to\infty} \frac{1}{T} \int_0^T {\rm tr}(\sigma_z \rho_t) \rd t . 
\label{eq:17} 
\end{eqnarray}
Note that $m$ is independent of the choice of the initial condition $\rho_0$. 
In figure~\ref{fig:2.1} we plot $m$ as a function of $\gamma^{-1}$, showing the 
characteristic behaviour of the order parameter in a second-order phase transition. 

Perhaps a surprising feature of the foregoing analysis is the observation that in the 
broken phase where eigenvalues are complex, convex combinations of energy 
eigenstates are not stationary. On account of the fact that the dynamical equation is 
autonomous, this implies, in particular, that if an initial state $\rho_0$ is chosen 
randomly out of the eigenstates $|\phi_j\rangle\langle\phi_j|$, each one of which 
is stationary, and the system evolves under the presence of gain and loss such that 
PT symmetry is broken, then after a passage of time the statistics obtained from 
$\rho_t$ are different from those obtained from $\rho_0$. This feature has no analogue 
in the standard unitary theory, and can be used to test the applicability of the model 
(\ref{eq:1}) in laboratory experiments, provided that a coherent control of the system, 
in such a way that gain and loss can be balanced without perturbing the system, 
is possible.  

If, on the other hand, a coherent implementation of gain and loss is not feasible, either 
because of fundamental quantum limits or current technological limits, then it is 
important to take into account additional effects arising from random perturbations. For 
this purpose, we shall assume that the 
model (\ref{eq:1}) remains valid, but in addition the system is perturbed at random. 
Specifically, we assume that the state is 
perturbed by Gaussian white noise in every orthogonal direction in the space of pure 
states, with strength $\sqrt{\kappa}$, where $\kappa\geq0$. In Ref.~\cite{BH2} it was 
shown that such a perturbation, when averaged out, generates a flow in the space of 
density matrices given by $\kappa \left( {\mathds 1}-n \rho \right)$; this, in turn, leads 
to the extended model: 
\begin{eqnarray}
\frac{\rd\rho}{\rd t} &=& -\ri [H,\rho] - \big( [{\mathit\Gamma},\rho]_+ - 2\, {\rm tr}
(\rho{\mathit\Gamma}) \rho \big) \nonumber \\ && 
+ \kappa \left( {\mathds 1}-n \rho \right), 
\label{eq:18}
\end{eqnarray}
where $\kappa\geq0$ and $n$ is the Hilbert space dimensionality. 

We proceed to analyse properties of (\ref{eq:18}). To begin, it should 
be evident that: (i) The evolution equation (\ref{eq:18}) preserves the overall probability so 
that ${\rm tr}(\rho_t)=1$ for all $t\geq0$. On account of the presence of noise, however, 
an initially pure state ceases to remain pure. In particular, we have: (ii) The evolution of 
the purity is governed by the equation 
\begin{eqnarray}
\frac{\rd}{\rd t}\, {\rm tr}\,\rho^2 &=& -4 \big( {\rm tr}({\mathit\Gamma}\rho^2)  
- {\rm tr}(\rho{\mathit\Gamma}) {\rm tr}(\rho^2) \big) \nonumber \\ && + 2\kappa 
\big( 1-n\, {\rm tr}(\rho^2) \big). 
\label{eq:19} 
\end{eqnarray}
Notice that when $\rho^2=\rho$ the first term in the right side of (\ref{eq:19}) vanishes, 
whereas the second term is negative. Hence an initially pure state necessarily evolves 
into a mixed state due to noise. (iii) When ${\mathit\Gamma}=0$, the solution to the 
dynamical equation (\ref{eq:18}) takes the form 
\begin{eqnarray}
\rho_t = \frac{1}{n}\left[ {\mathds 1} + \left(n\, \re^{-{\rm i}Ht} \rho_0 \re^{{\rm i}Ht} 
- {\mathds 1} \right) \re^{-\kappa nt} \right], 
\label{eq:20}
\end{eqnarray}
and has a single fixed point $\rho_\infty=n^{-1}{\mathds 1}$. This can be seen from 
the facts that the off diagonal elements of $\kappa \left( {\mathds 1}-n \rho \right)$ are 
negative, and that the diagonal elements are positive (\textit{reps}. negative) if the 
diagonal element of $\rho$ is less than (\textit{reps}. larger than) $n^{-1}$. 
The effect of the term $\kappa \left( {\mathds 1}-n \rho \right)$ therefore is to generate a 
gradient flow towards the uniformly mixed state $\rho_\infty=n^{-1}{\mathds 1}$. (iv) Like the 
model (\ref{eq:1}), the evolution equation (\ref{eq:18}) is positive and autonomous in the 
sense described above. (v) The dynamical equation 
satisfied by an observable $\langle F\rangle={\rm tr}(F \rho_t)$ is given by 
\begin{eqnarray}
\frac{\rd \langle F\rangle}{\rd t} &=& \ri \langle[H,F]\rangle - \langle[{\mathit\Gamma},F]_+
\rangle + 2\, \langle{\mathit\Gamma}\rangle \langle F\rangle \nonumber \\ && + \kappa 
\big( {\rm tr}(F)-n\langle F\rangle \big) .  
\label{eq:21}
\end{eqnarray}

To gain insights into fixed-point structures of the extended model (\ref{eq:18}) we have 
performed numerical studies based on a two-level system with the Hamiltonian 
$K=\sigma_x - \ri \gamma \sigma_z$. Our analysis shows that once the noise 
strength $\kappa$ is turned on, irrespective of its magnitude, the phase transition is 
suppressed. This can be seen by the consideration of the order parameter $m$. As 
indicated above, in the noise-free case there is a clear indication of a second-order 
phase transition at the critical point $\gamma_c=1$. Under a noisy environment, 
however, the symmetry is broken for all values of $\gamma\neq0$; instead, for each 
value of $\gamma,\kappa$ an equilibrium state $\rho^*$ is established, to which 
every initial state converges. In figure~\ref{fig:2.2} we plot $m$ as a function of 
$\gamma^{-1}$ for a range of values for $\kappa>0$, showing the removal of the 
phase transition, although for sufficiently small $\kappa$ the signature of the transition 
is visible. We have performed further numerical and analytical studies of the evolution 
equation (\ref{eq:18}), the results of which indicate that the emergence of a nontrivial 
equilibrium state is a generic feature of the model. 

In summary, we have introduced a model for describing the evolution of a density 
matrix for a system having gain and loss, and investigated its properties in detail. 
In particular, we have identified the associated stationary states, and shown the 
existence of phase transitions in the generic mixed-state context (cf. 
figure~\ref{fig:2.1}). We then extended the model to include a noise term, and showed 
evidences for the existence of equilibrium states that eliminate phase transitions 
(cf. figure~\ref{fig:2.2}). The applicability of our models is expected to be verifiable 
in laboratory experiments.

\vspace{0.2cm} 
EMG acknowledges support via the Imperial College JRF scheme. 
We thank H. F. Jones for comments.

\end{document}